\documentclass{iopart}
\usepackage{graphicx}
\renewcommand{\a}{a}
\newcommand{\avec}{\mbox{\boldmath$\a$}}
\newcommand{\lh}{l_{h_0}}
\newcommand{\hvec}{\mbox{\boldmath$d$}}
\begin{document}
\title{Detecting gravitational radiation from neutron stars using
a six-parameter adaptive MCMC method}


\author{Richard Umst\"atter$^1$\footnote{richard@stat.auckland.ac.nz},
Renate Meyer$^1$\footnote{meyer@stat.auckland.ac.nz},
R\'ejean J. Dupuis$^2$\footnote{rejean@astro.gla.ac.uk},
John Veitch$^2$\footnote{jveitch@astro.gla.ac.uk},
Graham Woan$^2$\footnote{graham@astro.gla.ac.uk}
and
Nelson Christensen$^3$\footnote{nchriste@carleton.edu}}
\address{$^1$Department of Statistics, University of Auckland,
Auckland, New Zealand
$^2$Department of Physics and Astronomy, University of Glasgow,
G12 8QQ, United Kingdom \\
$^3$Physics and Astronomy, Carleton College,
Northfield, MN 55057, USA\\
}
\date{\today}
\begin{abstract}
We present a Markov chain Monte Carlo technique for detecting
gravitational radiation from a neutron star in laser
interferometer data.  The algorithm can estimate up to six unknown
parameters of the target, including the rotation frequency and
frequency derivative, using reparametrization, delayed rejection
and simulated annealing. We highlight how a simple extension of
the method, distributed over multiple computer processors, will
allow for a search over a narrow frequency band. The ultimate goal
of this research is to search for sources at a known
locations, but uncertain spin parameters, such as may be found in SN1987A.
\end{abstract}
\section{Introduction}
Rapidly rotating neutron stars could be an important source
of gravitational wave signals. Several mechanisms have been
proposed that would cause them to emit quasi-periodic
gravitational waves~\cite{Cutler02,bild}.

Interferometric gravitational wave detectors are now operating in
numerous locations around the world~\cite{GEO, LIGO, VIRGO, TAMA},
and much work has gone into the development of dedicated search
algorithms for these signals. Radio observations can provide the
sky location, rotation frequency and spin-down rate of known
pulsars, and this knowledge simplifies the analysis. This was the
case for the recent search for a signal from
PSR~J1939+2134~\cite{LIGO-CW}. When the position and phase
evolution of a source are not known, all-sky hierarchical
strategies are required, and these have huge computational
requirements~\cite{jaranowski,stack}.

Here we concentrate on the search for a gravitational wave signal
from a known location, but where spin parameters of the rotating
neutron star are not well known (but within a narrow band). SN1987A is a good example of a
poorly parameterised source for which the sky location is
known, but where there are large uncertainties in
the frequency and spin-down parameters of a putative neutron
star~\cite{midd}. In particular, we consider a search with six
unknown parameters: the gravitational wave amplitude $h_0$, the
polarization angle $\psi$ (which depends on the position angle of
the spin axis in the plane of the sky), the phase of the signal at
a fiducial time $\phi_0$, the inclination of the spin axis with
respect to the line-of-sight $\iota$, the uncertainty in the
absolute value of the signal frequency $\Delta f$, and the
frequency derivative $\Delta\dot{f}$.

We use a Bayesian Markov chain Monte Carlo (MCMC) technique for
this analysis as MCMC methods have been applied successfully to
similar problems involving large numbers of parameters
\cite{gilks96}. In a previous study \cite{nlc1}, we used a
Metropolis-Hastings (MH) algorithm \cite{metr53,hastings} for a
similar search, but with only five parameters ($\Delta\dot{f}$
being absent). When the frequency derivative $\Delta\dot{f}$ is
included in the basic MH routine of~\cite{nlc1} the large
correlation between $\Delta f$ and $\Delta\dot{f}$ makes the
parameter search difficult, and the basic MH algorithm becomes
inefficient. In order to adequately sample the parameter space we
implemented a combination of three different strategies for
accelerating convergence of Markov chains: reparameterisation, the
delayed rejection method of Tierney and Mira~\cite{TierneyMira99}
(which is an adaptive version of the MH algorithm), and simulated
annealing \cite{Kirk83} (which is a Monte Carlo approach to global
optimization). The parameter $\Delta f$ is highly correlated with
$\Delta\dot{f}$, and a strong correlation also exists between
$h_0$ and $\cos\iota$. An initial transformation of these
variables to near orthogonality yields a more tractable parameter
space that is more effectively sampled.

The heterodyne manipulation of the data used in this study is
identical to that presented (by two of us) in an end-to-end robust
Bayesian method of searching for periodic signals in gravitational
wave interferometer data~\cite{DupuisWoan03}, and is also
described in~\cite{LIGO-CW}. A brief summary of this heterodyne
technique is given in Sec.~\ref{signal}. Our delayed rejection
method, as well as the reparameterisation strategy, is presented
in Sec.~\ref{MH}. In Sec.~\ref{results} we present the results of
this study, using  synthesized signals, for this six parameter
problem. A brief discussion of the long term goals for this work
are presented in Sec.~\ref{disc}.

\section{The gravitational wave signal}
\label{signal}

Gravitational waves from spinning neutron stars are expected to be
weak at the Earth, therefore long integration periods are
necessary to extract the signal.  It is therefore important to
take proper account of the antenna patterns of the detectors and
the Doppler shift due to the motion of the Earth.

As in previous studies \cite{LIGO-CW,nlc1,DupuisWoan03} we
consider the signal expected from a non-precessing triaxial
neutron star. The gravitational wave signal from such an object is
at twice its rotation frequency, $f_{\rm s}=2 f_{\rm r}$, and we
characterise the amplitudes of each polarization with overall
strain factor, $h_0$. The measured gravitational wave signal will
also depend on the antenna patterns of the detector for the
`cross' and `plus' polarisations, $F_{\times, +}$, giving a signal
\begin{equation}
s(t) = \frac{1}{2}F_{+}(t;\psi)h_{0}(1 + \cos^{2}\iota)\cos
\Psi(t) + F_{\times} (t;\psi)h_{0}\cos \iota \sin  \Psi(t),
\label{s}
\end{equation}
A simple slowdown model provides the phase evolution of the signal
as
\begin{equation}
 \Psi(t) = \phi_{0} + 2\pi \left[f_{\rm s}(T - T_{0}) + \frac{1}{2}\dot{f_{\rm s}} (T -
       T_{0})^{2} 
       \right],
\label{phase1}
\end{equation}
where
\begin{equation}
T = t + \delta t= t + \frac{\vec{r} \cdot \vec{n}}{c}  +
\Delta{T}. \label{time}
\end{equation}
Here, $T$ is the time of arrival of the signal at the solar system
barycenter, $\phi_{0}$ is the phase of the signal at a fiducial
time $T_{0}$, $\vec{r}$ is the position of the detector with
respect to the solar system barycenter, $\vec{n}$ is a unit vector
in the direction of the neutron star, $c$ is the speed of light,
and $\Delta{T}$ contains the relativistic corrections to the
arrival time \cite{Taylor}.

If $f_{\rm s}$ and $\dot{f}_{\rm s}$ are known from (for example)
radio observations, the signal can be \emph{heterodyned} by
multiplying the data by $\exp[-i\Psi(t)]$, low-pass filtered and
resampled, so that the only time varying quantity remaining is the
antenna pattern of the interferometer. We are left with a simple
model with four unknown parameters $h_0$, $\psi$, $\phi_{0}$ and
$\iota$.  If there is an uncertainty in the frequency and frequency derivative then 
we have two additional parameters, the differences
between the signal and heterodyne frequency and frequency
derivatives, $\Delta f$ and $\Delta\dot{f}$, giving a total of six
unknown parameters.

A detailed description of the heterodyning procedure is presented
elsewhere~\cite{LIGO-CW,DupuisWoan03}.  Here we just provide a
brief summary of this standard technique. The raw signal, $s(t)$,
is centered near twice the rotation frequency of the neutron star,
but is Doppler modulated due to the motion of the Earth and the
orbit of the neutron star if it is in a binary system. The
modulation bandwidth is typically $~10^4$ times less than the
detector bandwidth, so one can greatly reduce the effective data
rate by extracting this band and shifting it to zero frequency. In
its standard form the result is one binned data point, $B_{k}$,
every minute, containing all the relevant information from the
original time series but at only $2\times10^{-6}$ the original
data rate. If the phase evolution has been correctly accounted for
at this heterodyning stage then the only time-varying component
left in the signal will be the effect of the antenna pattern of
the interferometer, as its geometry with respect to the neutron
star varies with Earth rotation. Any small error, $\Delta f$, in
the heterodyne frequency will cause the signal to oscillate at
$\Delta f$ (plus the residual Dopper shift). We estimate the noise variance, $\sigma^2_k$, in the
bin values, $B_{k}$, from the sample variance of the contributing
data. It is assumed that the noise is stationary over the 60\,s of
data contributing to each bin.

\section{The adaptive Metropolis-Hastings algorithm}
\label{MH} After heterodyning, the signal on which we wish to
carry out our MCMC analysis has the form~\cite{DupuisWoan03}
\begin{equation}
\fl y(t_k;\avec)  =\frac{1}{4}F_+(t_k;\psi)h_{0} (1 +
 \cos^2\iota)e^{i\Delta\Psi(t)}
               - \frac{i}{2}F_\times(t_k;\psi) h_{0} \cos\iota
                 e^{i\Delta\Psi(t)},\label{yeq}
\end{equation}
where $t_k$ is the time of the $k^\textrm{th}$ bin $B_k$ and
$\avec = (h_0,\,\psi,\,\phi_0,\,\cos\iota,\,\Delta
f,\,\Delta\dot{f})$ is a vector of our unknown parameters.
$\Delta\Psi(t)$ represents the residual phase evolution of the
signal, equalling $\phi_0+2\pi[\Delta
f(T-T_0)+\Delta\dot{f}(T-T_0)^{2}/2]$. The objective is to fit this
model to the antenna output data
\begin{equation}
 B_k=y(t_k;\avec) +\epsilon_k,
\end{equation}
where $\epsilon_k$ is assumed to be normally distributed noise
with a mean of zero and known variance $\sigma_k^2$. Assuming
exchangeability of the binned data points, $B_k$, the joint
likelihood that these data ${\bf d}=\{B_k\}$ arise from a model
with a certain parameter vector $\avec$ is \cite{DupuisWoan03}
\begin{equation}
 p({\bf d}|{\avec})\propto\prod_k \exp\left[-\frac{1}{2}\left| \frac{B_k-y(t_k;{\avec})}{\sigma_k}\right| ^2 \right]
   = \exp \left[ \frac{-\chi^2(\avec)}{2}\right],
\end{equation}
where
\begin{equation}
 \chi^2(\avec)=\sum_k{ \left| \frac{B_k-y(t_k;\avec)}{\sigma_k} \right| ^2}.
\end{equation}
In order to draw any inference on the unknown parameter vector
$\avec$ we need the (posterior) probability of $\avec$ given
$\hvec$ which can be obtained from the likelihood via an
application of Bayes' theorem. The unnormalized posterior density
\begin{equation}
p(\avec|\hvec)\propto p(\avec)p(\hvec|\avec)
\label{eq:post}
\end{equation}
is the product of the prior density of $\avec$, $p(\avec)$, and
the joint likelihood. Accordingly, appropriate priors have to be
chosen for the particular parameters. In this study we use uniform
priors with prior ranges $[-\pi,\pi]$, $[-\pi/4,\pi/4]$ and
$[-1,1]$ for the angle parameters $\phi_0$, $\psi$ and $\cos
\iota$ respectively. For $h_0$ we also specify a uniform prior
with boundary $[0,1000]$ in units of the rms noise
\cite{DupuisWoan03}. For the frequency and spindown uncertainty 
we use suitable uniform priors with ranges of
$[-\frac{1}{60},\frac{1}{60}]$\,Hz  and
$[-10^{-9},10^{-9}]$\,Hz\,s$^{-1}$ for $\Delta f$ and
$\Delta\dot{f}$, respectively.

The normalized posterior density $p(\avec|\hvec)=
p(\avec)p(\hvec|\avec)/p(\hvec)$ cannot be evaluated analytically,
so we use Monte Carlo methods to explore $p(\avec|\hvec)$. If we
can simulate from $p(\avec|\hvec)$, we can estimate all
interesting quantities, including the posterior means of all
parameters from the corresponding sample means, to any desired
accuracy by increasing the sample size.

However, drawing independent samples in a six-dimensional
parameter space is not feasible. It has already been shown that
MCMC methods can be used to parameterise gravitational wave
signals of low signal-to-noise ratio~\cite{nlc1} with four unknown
parameters. These simulate a Markov chain, constructed so that its
stationary distribution coincides with the posterior distribution
and the sample path averages converge to the expectations. A
minimal requirement for this is the irreducibility of the chain
and hence the ability of the chain to reach all parts of the state
space ~\cite{gilks96}. A specific MCMC technique is the MH
algorithm~\cite{metr53,hastings} which does not require the
normalization constant, only the unnormalized posterior density
of Eq.~(\ref{eq:post}). We employed the MH algorithm for the four
and five parameter pulsar detection problems~\cite{nlc1}. The
efficiency of the MH algorithm depends heavily on the choice of
the proposal density. Intuition suggests that the closer the
proposal distribution is to the target, the faster convergence to
stationarity is achieved. Default choices such as a Gaussian
proposal or a random walk result in very slow mixing for this
6-parameter problem. To increase the speed of convergence, we
employed an \emph{adaptive} technique, adaptive in the sense that
it allows the choice of proposal distribution to depend upon
information gained from the already sampled states as well as the
proposed but rejected states. The idea behind the delayed
rejection algorithm specified by~\cite{TierneyMira99} is that
persistent rejection, perhaps in particular parts of the state
space, may indicate that locally the proposal distribution is
badly calibrated to the target. Therefore, the MH algorithm is
modified so that on rejection, a second attempt to move is made
with a proposal distribution that depends on the previously
rejected state. This adaptive Monte-Carlo
method~\cite{TierneyMira99} was generalized for the variable
dimension case~\cite{GreenMira01} and renamed the `delayed
rejection method'. Since we have a fixed dimension problem here
we implemented the original version~\cite{TierneyMira99}, and
also the generalization~\cite{GreenMira01} that uses the
reversible jump method. It turned out that the delayed rejection
with the reversible jump method was not that beneficial for  this
particular problem and thus we will explain the original delayed
rejection algorithm~\cite{TierneyMira99} here.

\begin{figure}[h]
  \begin{center}
    \includegraphics[width=1.0\textwidth]{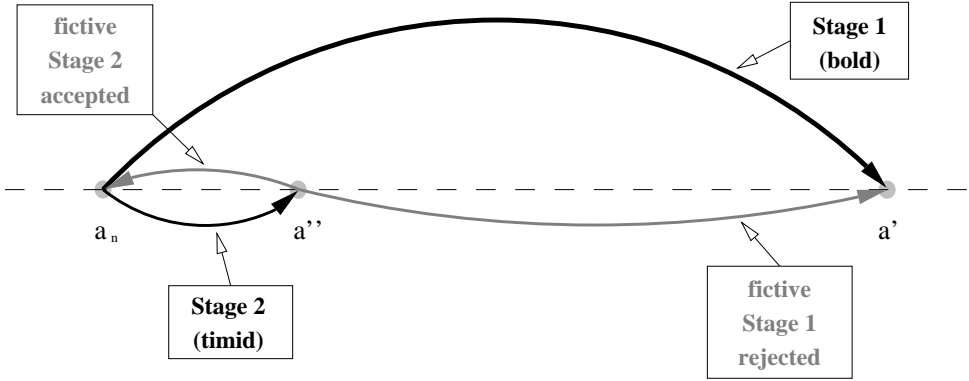}
  \end{center}
\caption{The delayed rejection method. In case of rejection of
the first, bold step a second, more timid move is proposed. In
order to maintain the reversibility of the Markov Chain the
acceptance probability has to consider a fictive return path.}
\label{fig1}
\end{figure}

For the Metropolis-Hastings algorithm a new state in a Markov
chain is chosen first by sampling a candidate $\avec'$ from a
certain proposal distribution $q_1(\avec'|\avec_n)$ usually
depending on the current state $\avec_n$ and then accepting or
rejecting it with a probability $\alpha_1(\avec'|\avec_n)$
depending on the distribution of interest. This rejection is
essential for the convergence of the chain to the intended target
distribution. The choice of a good proposal distribution is
important to avoid persistent rejections in order to achieve good
convergence of a chain. However in different parts of the state
space different proposals are required. When a proposed MH move is
rejected, a second candidate $\mbox{\boldmath$\a$}''$ can be
sampled with a different proposal distribution
$q_2(\avec''|\avec',\avec_n)$ that can depend on the previously
rejected proposal. Since a rejection suggests a bad fit of the
first proposal, a different form of proposal can be advantageous
in the second stage. To preserve reversibility of the Markov chain
and thus to comply with the detailed balance condition, the
acceptance probabilities for both the first and the second stage
are given by \cite{Mira98}
\begin{equation}
\fl \alpha_1(\avec '|\avec_n) =
 \min\left(1,\frac{p(\avec')p(\hvec|\avec')
 q_1(\avec_n|\avec')}{p(\avec_n)p(\hvec|\avec_n)q_1(\avec'|\avec_n)}\right)
\end{equation}
and
\begin{equation}
\fl \alpha_2 (\avec''|\avec_n) = \min\left(1,\frac
{p(\avec'')p(\hvec|\avec'') q_1(\avec'|\avec'')
q_2(\avec_n|\avec',\avec'')[1
-\alpha_1(\avec'|\avec'')]}{p(\avec_n)p(\hvec|\avec_n)
q_1(\avec'|\avec_n)
q_2(\avec''|\avec_n,\avec')[1-\alpha_1(\avec'|\avec_n)]}\right)
\end{equation}
respectively. Fig.~\ref{fig1} illustrates the idea of delayed
rejection. When the second stage proposal step is applied due to
rejection of the first, the chain has, in order to preserve the
reversibility, to imply a return path which comprises a fictive
stationary Markov chain consisting of a fictive stage 1 proposal
step from $\avec''$ to $\avec'$ which is rejected followed by an
accepted fictive second stage move to
$\avec_n$~\cite{GreenMira01}.
Although the delayed rejection method provides better acceptance
rates over the two stages, cross-correlations between the
parameters still impede convergence of the Markov chain.
Preliminary runs reveal that especially the parameters $\Delta f$
and $\Delta\dot{f}$, and to a certain extent $h_0$ and $\cos\iota$
are highly correlated after the Markov chain has found
a potential mode. The consequence of which is poor mixing of the
chain and therefore a reparameterisation is required.

The coherence between $\Delta f$ and $\Delta\dot{f}$ is obvious
since the data is sampled from time $t_{\rm start}$ to $t_{\rm
end}$, where the heterodyned signal traverses a frequency from
$f_{\rm start}={\Delta f}+\frac{1}{2} \Delta\dot{f} \cdot t_{\rm
start}$ to $f_{\rm end}={\Delta f}+\frac{1}{2} \Delta\dot{f} \cdot
t_{\rm end}$; time $t=0$ is an epoch time when $f=\Delta f$. 
Hence it is much more natural to work with these
frequencies as new parameters and vary them with a certain
correlation which influences $\Delta\dot{f}$ indirectly. The
original parameters are then obtained by the simple linear
transformation
\begin{equation}
{\Delta f}=f_{\rm start}-\frac{1}{2} \Delta\dot{f} \cdot t_{\rm
start}
\end{equation}
and
\begin{equation}
\Delta\dot{f}=2 \cdot \frac{f_{\rm end}-f_{\rm start}}{t_{\rm
end}-t_{\rm start}}.
\end{equation}
Since the Jacobian of this transformation is constant the prior
distributions for the new parameters $f_{\rm start}$ and $f_{\rm
end}$ are flat as well.

\begin{figure}[h]
  \begin{center}
    \includegraphics[width=9.4234cm]{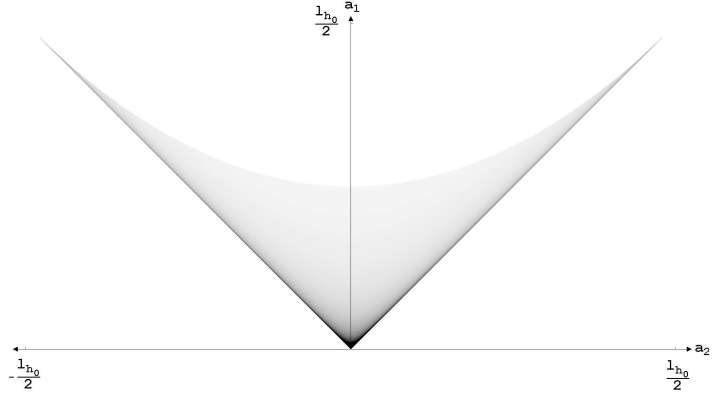}
  \end{center}
\caption{Joint prior density of $a_1$ and $a_2$ for a given
boundary $\lh$ for the parameter $h_0$} \label{fig2}
\end{figure}

Another cross-correlation can be observed between the parameters
$h_0$ and $\cos \iota$ that arises from the fact that $h_0$ can be
seen as a scaling factor and $\cos\iota$ as a non-linear weighting
between the plus and cross polarisation part of the model. As seen in Eq.
\ref{yeq}, the plus part is multiplied by the factor
$a_1=\frac{1}{4}h_0(1+\cos^2\iota)$ while the cross part
encloses the term $a_2=\frac{1}{2}h_0 \cos \iota $. The original
parameters can be derived from
\begin{equation}
h_0=2 \left( a_1 + \sqrt{a_1^2-a_2^2} \right),
\end{equation}
and
\begin{equation}
\cos \iota =\frac{2a_2}{h_0}.
\end{equation}
As mentioned above, the prior distribution of the parameters $h_0$
and $c=\cos\iota$ are chosen uniform with joint probability
density function
\begin{equation}
 f(h_0,c)=\left\{
\begin{array}{ll}
(2 \lh)^{-1}, & \textrm{if } 0 \le h_0 < \lh \textrm{, } -1 \le c
\le 1,
\\ 0, & \textrm{otherwise,}  \\
\end{array}
\right.
\end{equation}
where for this study $l_{h_0}=1000$ in units of the rms noise.
This implies a joint prior distribution for the parameters $a_1$
and $a_2$ of the form
\begin{equation}
g(a_1,a_2)  =  \left\{
\begin{array}{ll}
(2 \lh)^{-1}, & \textrm{if } \left| a_2 \right| \le a_1 < \frac{4a_2^2+\lh^2}{4 \lh}  \le \frac{\lh}{2}
\\ 0, & \textrm{otherwise} \\
\end{array}
 \right\} \left| \det J \right| \nonumber \\
\end{equation}
with Jacobian
\begin{equation}
\det J = \frac{2}{\sqrt{a_1^2-a_2^2}}.
\end{equation}
Since the Jacobian is positive for the above restrictions we can
write
\begin{equation}
g(a_1,a_2) = \left\{
\begin{array}{ll}
\frac{1}{\lh \sqrt{a_1^2-a_2^2}}, &\textrm{if } \left| a_2 \right|  \le\  a_1
 < \frac{4 a_2^2+\lh^2}{4 \lh} \le \frac{\lh}{2},
\\ 0, & \textrm{otherwise.} \\
\end{array}
\right.
\end{equation}
This joint prior density has the shape shown in Fig.~\ref{fig2}.
These reparameterisations result in a fast mixing Markov chain
but still, the choice of a suitable proposal distribution is
essential. Usually, a multivariate Normal distribution is utilized
for the proposal distributions $q_1(\avec'|\avec_n)$ and
$q_2(\avec''|\avec',\avec_n)$, with means equal to the current
state and different variances depending on the stage. Larger
variances are chosen for the `bold' first stage steps, while
smaller variances are more beneficial for the `timid' second stage
candidates. The covariance matrix has to comprise the correlation
between the parameters $f_{\rm start}$ and $f_{\rm end}$ since
this correlation indirectly controls the parameter
$\Delta\dot{f}$ as mentioned above. Hence choosing proposals for
$f_{\rm start}$ and $f_{\rm end}$ with a correlation of $1$ would
imply no change of $\Delta\dot{f}$ because both parameters are
changed in the same way while a correlation of $0$ would have a
great impact on $\Delta\dot{f}$ since $f_{\rm start}$ and $f_{\rm
end}$ are changed completely uncorrelated. Thus the correlation
between $f_{\rm start}$ and $f_{\rm end}$ has to be treated
randomly in order to control $\Delta\dot{f}$. Best results are
obtained when a correlation of $0$ is chosen with probability
$0.5$ for the bold moves of $\Delta\dot{f}$ and a correlation of
almost $1$ otherwise for timid moves of $\Delta\dot{f}$. The
proposals for the parameters $a_1$ and $a_2$ are sampled
independently since they represent scaling factors for the plus
and cross polarisation part, respectively. Finally we have to consider the
correlation between the original parameters $\psi$ and $\phi_0$
which  are not reparameterised. Pilot runs show that they are
highly correlated. Hence the proposal
distribution is adapted accordingly.

Unfortunately, the posterior distribution features very narrow
modes in a large parameter space that has to be scanned. Thus a
simple Normal distribution is not suitable for a proposal
distribution as pilot runs have revealed. Instead, a proposal
distribution with long tails and strong narrow mode is required.
This can easily be achieved by generating a random sample between
two boundaries $b_l$ and $b_h$ for the standard deviation of the
proposal by generating a random weight for the weighted geometric
mean of these two boundaries. Hence we sample standard deviations
according to $\sigma=b_h^w b_l^{1-w}$ where $w \sim \beta(a,b)$ is
Beta-distributed with parameters $a$ and $b$. The resulting
proposal distribution is symmetric with very long tails and a
strong narrow mode. In order to obtain higher standard deviations
for the first stage the choice of $w \sim \beta(2,1)$ (with mean
$\frac{2}{3}$) is adequate while for the second stage $w \sim
\beta(1,2)$ (with mean $1/3$) samples smaller standard deviations.

The implementation of the ideas outlined above leads to reasonable
acceptance rates and hence to a much better convergence of the
Markov chain. While during the burn-in period it is mainly the
stage 1 candidates that are accepted, the Markov chain is driven
mainly by stage 2 candidates after the burn-in. But still, the
stationary distribution features many distinct modes that carry
the risk of  trapping the Markov chain. Therefore, we regard the
posterior as a canonical distribution
\begin{eqnarray}
  p({\avec}|\hvec) &\propto p(\avec)p(\hvec|\avec)\propto p(\avec)\exp\left[-\frac{\chi^2(\avec)}{2} \right] \nonumber \\
  &\propto\exp\left[-\frac{\chi^2(\avec)-2\log \left[ p(\avec) \right]}{2} \right] \nonumber \\
  &\propto\exp\left[-\beta\left(\chi^2(\avec)-2\log\left[p(\avec)\right]\right)\right].
\end{eqnarray}
with inverse temperature $\beta$. During the burn-in period this
inverse temperature can pass through values starting at a low
value (thus high temperature) and ending up at $\beta=\frac{1}{2}$
which coincides with the posterior distribution. This simulated
annealing technique was introduced by Metropolis et
al.~\cite{metr53} and allows scanning of the whole parameter space
by permitting larger steps. For the annealing schedule an
exponential temperature curve is applied. For a certain number of
iterations $t_{s}$, it starts with an inverse temperature
$\beta_0$ until it reaches $\beta=\frac{1}{2}$. The inverse
temperature follows the function
\begin{equation}
 \beta(t)= \left\{
\begin{array}{ll}
 \beta_0 \exp \left[ \frac{t}{t_{s}} \log \left[ \frac{\beta}{\beta_0} \right]
 \right], & \textrm{if } 0 \leq t \leq t_s, \\
 \frac{1}{2}, & \textrm{if } t > t_s,
 \end{array} \right.
\end{equation}
depending on the current iteration $t$. Since the starting
temperature is dependent on the data set which is influenced by
the amplitude $h_0$ of the signal it has to be adapted
accordingly.

\section{Results with simulated signals}
\label{results} We have synthesized fictitious data, and passed it
through our six parameter MCMC routine. The presentation of
results here is similar to that of the four and five parameter
study of~\cite{nlc1}. The artificial signals were embedded within
white and normally distributed noise. The ability of the MCMC
algorithm to successfully find the signal and estimate the six
parameters was demonstrated, and is presented below. The
artificial signals $s(t)$ were synthesized assuming a source at RA
$=4^{\rm h}\,41^{\rm m}\,54^{\rm s}$ and dec
$=18^\circ\,23'\,32''$, as would be seen by the LIGO-Hanford
interferometer. The signals were then added to noise; we assumed a
signal at $300$\,Hz and a corresponding noise spectral density of
at that frequency of $h(f)=8\times 10^{-23}$\,Hz$^{-1/2}$. The
amplitude of the signal used in our test runs was varied in the
range $h_0=4.0\times 10^{-24}$ to $1.5\times 10^{-22}$. The length
of the data set corresponded to $14\,400$ samples or 10 days of
data at a rate of one sample per minute (which was the rate used
for the LIGO/GEO S1 analysis described in~\cite{LIGO-CW}).

\begin{figure}[h]
  \begin{center}
    \includegraphics[width=9.4234cm,angle=0]{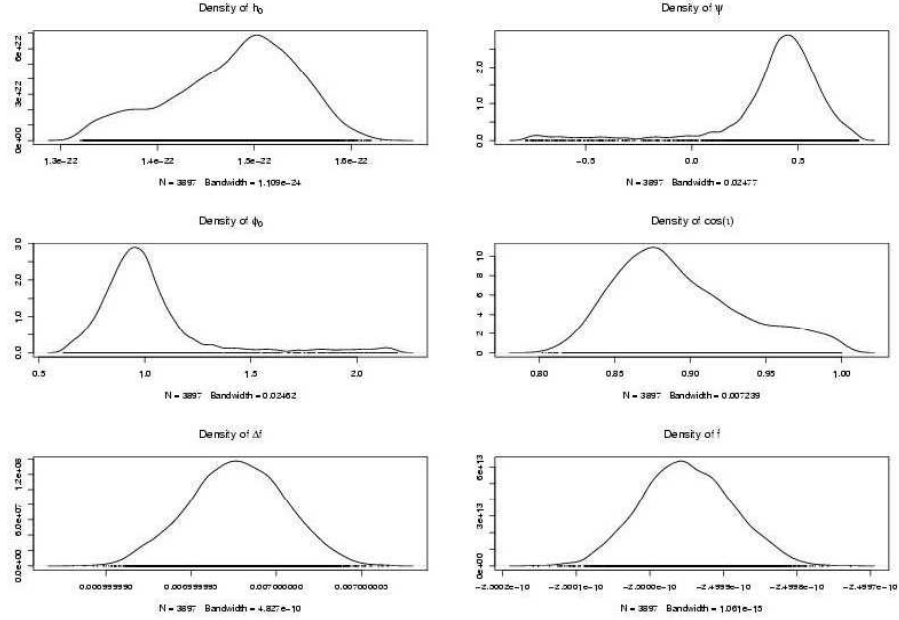}
  \end{center}
\caption{MCMC estimates of the posterior pdf (kernel density) for
the six parameters $h_0$, $\psi$, $\phi_{0}$, $cos\iota$, $\Delta
f$ and $\Delta\dot{f}$. This synthesized signal had {\it real}
parameters of: $h_0=15\times 10^{-23}$, $\psi=0.4$,
$\phi_{0}=1.0$, $cos\iota=0.878$, $\Delta f=7.0\times 10^{-3} Hz$
and $\Delta\dot{f}=-2.5\times 10^{-10}$. The mean of the $h_0$
distribution here is $14.84\times 10^{-23}$.} \label{fig3}
\end{figure}

In Fig.~\ref{fig3} we display the MCMC generated posterior
probability distribution functions (pdfs) for an example signal.
The \emph{real} parameters for this signal were: $h_0=1.5\times
10^{-22}$, $\psi=0.4$, $\phi_{0}=1.0$ (both in radians),
$\cos\iota=0.878$, $\Delta f=7.0\times 10^{-3}$\,Hz and
$\Delta\dot{f}=-2.5\times 10^{-10}$\,Hz\,s$^{-1}$. For this
example the program ran for $10^6$ iterations. For a signal this
large only about $2.5 \times 10^4$ iterations were needed for the
burn-in, and this data is discarded from the analysis. Short-term
correlations in the chain were eliminated by `thinning' the
remaining terms; we kept every $250^{\rm th}$ item in the chain.
In this example the MCMC yielded median values and 95\% posterior
probability intervals of $h_0=14.91\times10^{-23}$ ($13.41\times
10^{-23}$ to $15.84\times 10^{-23}$), $\psi=0.439$ ($-0.552$
 to $0.707$), $\phi_0=0.964$ ($0.696$ to $1.958$), $\cos\iota
=0.884$ ($0.828$ to $0.988$), $\Delta f = 6.99999772 \times
10^{-3}$\,Hz ($6.99999217 \times 10^{-3}$\,Hz to $7.00000314
\times 10^{-3}$\,Hz) and $\Delta\dot{f} = -2.4999541 \times
10^{-10}$\,Hz\,s$^{-1}$ ($-2.5000767 \times
10^{-10}$\,Hz\,s$^{-1}$  to $-2.4998272 \times
10^{-10}$\,Hz\,s$^{-1}$). The 95\% posterior probability interval
is specified by the 2.5\% and 97.5\% quartiles of $p(
a_{i}|\hvec)$. 

With the noise level used, $h(f)=8\times 10^{-23}$, we were able
to successfully detect signals with amplitudes of $h_0 \ge
4.0\times 10^{-24}$ with 10 days of data. This should be compared
with the results presented in~\cite{nlc1} where with just four
parameters ($h_0$, $\psi$, $\phi_0$ and $cos\iota$), we were able
to confidently detect signals with an amplitude four times
smaller. The addition of the new frequency parameters has the
disadvantage of complicating the search due to the corresponding
increase in the size of the parameter space. For our study, we let
the initial burn-in of the Markov chain last for as long as
$3.5\times 10^5$ iterations, and if the signal was not found by
this time the search was terminated. It may be possible to find
smaller signals with a longer burn-in.

\begin{figure}[h]
  \begin{center}
    \includegraphics[width=4.8cm,angle=-90]{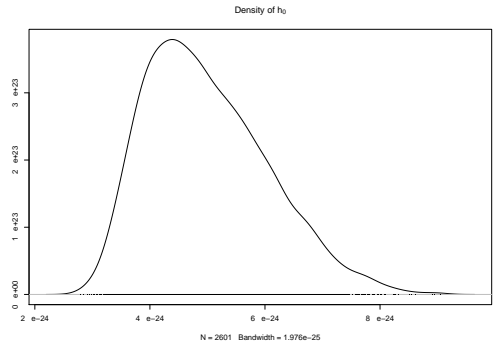}
  \end{center}
\caption{MCMC estimate of the posterior pdf (kernel density) for
the parameter $h_0$ from a six parameter search using synthesized
data. The real parameter value for this signal was $h_0=4.0\times
10^{-24}$. This was the smallest signal detectable by the MCMC
method for the noise level used.} \label{fig4}
\end{figure}

\begin{figure}[h]
  \begin{center}
    \includegraphics[width=4.8cm,angle=-90]{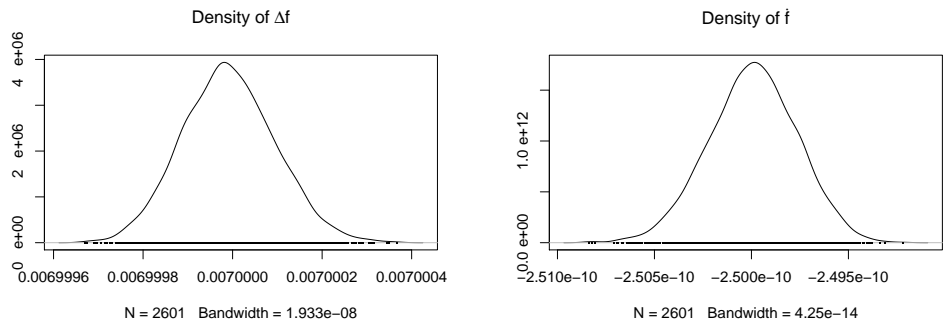}
  \end{center}
\caption{MCMC estimate of the posterior pdfs (kernel densities)
for the parameters $\Delta f$ and $\Delta\dot{f}$ from a six
parameter search using synthesized data with the smallest
detectable signal $h_0=4.0 \times 10^{-24}$. The real parameters
for this signal were: $\Delta f=7.0 \times 10^{-3}$\,Hz and
$\Delta\dot{f}=2.5 \times 10^{-10}$\,Hz,s$^{-1}$.} \label{fig5}
\end{figure}

Fig.~\ref{fig4} shows the MCMC estimated posterior for the
smallest value of the parameter $h_0$ that we were able to
identify with the MCMC code. The true parameter values for this
run were $h_0=4.0\times 10^{-24}$, $\psi=0.4$, $\phi_0=1.0$,
$\iota=0.5$ ($\cos\iota =0.878$), $\Delta f=7.0 \times
10^{-3}$\,Hz and $\Delta\dot{f} =-2.5 \times
10^{-10}$\,Hz\,s$^{-1}$. In this run the MCMC yielded a mean value
and 95\% posterior probability interval of $h_0=4.8\times
10^{-24}$ ($0.34\times 10^{-24}$ to $0.74\times 10^{-24}$).
Fig.~\ref{fig5} displays the MCMC estimated posterior for the
parameters $\Delta f$ and $\Delta\dot{f}$, which provides mean
values and 95\% posterior probability intervals of $\Delta f=7.0
\times 10^{-3}$\,Hz ($6.9998 \times 10^{-3}$\,Hz to $7.0002 \times
10^{-3}$\,Hz), and $\Delta\dot{f}=-2.500 \times
10^{-10}$\,Hz,s$^{-1}$ ($-2.505 \times 10^{-10}$\,Hz,s$^{-1}$ to
$-2.496 \times 10^{-10}$\,Hz,s$^{-1}$). As can be seen from
Figs.~\ref{fig4} and \ref{fig5}, even with small signal level it
is still possible to extract the most astrophysically important
parameters. For this MCMC run there were a total of $10^6$
iterations, with the first $3.5 \times 10^5$ as the burn-in.

\section{Discussion and conclusions}
\label{disc} In the simplest application, the method demonstrated
here could complement searches for signals from known
pulsars~\cite{LIGO-CW,DupuisWoan03}; our method could be used to
verify the frequency and frequency derivative values. The real
advantage of the technique would come about in a search for a
signal at a known location, but where the frequency information
pertaining to the neutron star is not well known; a search for a
signal from SN1987A~\cite{midd} would be a possible application.
In the demonstration here the heterodyning process provides a band
of $1/60$\,Hz. It would be straightforward to expand this search
to a bandwidth of $5$\,Hz by running the code on 300 processors, a
task easily accomplished on a cluster of computers. For 10 days of
data it takes a single 2.8\,GHz personal computer approximately an
hour to conduct about $3.3 \times 10^4$ iterations of our MCMC
code. There are more iterations done per time interval at the
beginning of a run because at that time more stage-1 steps are
accepted. We believe that these MCMC methods offer great potential
benefits for gravitational radiation searches where the signals
depend on a large number of parameters.

\verb''\ack This work was supported by the National Science
Foundation Grants PHY-0071327 and PHY-0244357, The Royal Society
of New Zealand Marsden Fund Grant UOA 204, the University of
Auckland Research Committee, the Natural Sciences and Engineering
Research Council of Canada, Universities UK, and the University of
Glasgow.


\Bibliography{99}
\bibitem{GEO} Willke B {\it et al} 2002 {\it Class. Quantum Grav.} {\bf{19(7)}} 1377
\bibitem{LIGO} Abramovici A,Althouse W E, Drever R W P, G\"ursel Y, Kawamura S, Raab F J,
Shoemaker D, Sievers L, Spero R E, Thorne K S, Vogt R E, Weiss R, Whitcomb S E and Zucker M E
1992 {\it Science} {\bf 256} 325
\bibitem{VIRGO} Caron B {\it et al} 1996 {\it Nucl. Phys. Suppl.} {\bf 48} 107
\bibitem{TAMA} Tsubono K 1997 {\it Gravitational Wave Detection} ed K Tsubono, M-K Fujimoto and K Kurodo
(Tokyo: Universal Academic) pp 183-91
\bibitem{Cutler02} Cutler C 2002 {\it Phys. Rev. D} {\bf 66} 084025
\bibitem{bild} Bildsten L 1998 {\it Astrophys. J.} {\bf 501} L89
\bibitem{LIGO-CW} Abbott B {\it et al} 2003 {\it Phys. Rev. D}
\bibitem{jaranowski} Jaranowski P, Kr\'olak A and Schutz B F 1998 {\it Phys. Rev. D} {\bf 58} 063001
\bibitem{stack} Brady P and Creighton T 2000 {\it Phys. Rev. D} {\bf 61} 082001
\bibitem{midd} Middleditch J, Kristan J A, Kunkel W E, Hill K M, Watson R D, Lucinio R,
Imamura J N, Steiman-Cameron T Y, Shearer A S, Butler R, Redfern M, Danks A C 2000
{\it New Astronomy} {\bf 5} 243
\bibitem{gilks96}  Gilks W R, Richardson S and Spiegelhalter D J 1996 {\it %
Markov Chain Monte Carlo in Practice} (London: Chapman and Hall)
\bibitem{metr53}  Metropolis N, Rosenbluth A W, Rosenbluth M N,
Teller A H, Teller E 1953 {\it J. Chem. Phys.} {\bf 21} 1087
\bibitem{hastings}  Hastings W K 1970 {\it Biometrika} {\bf 57} 97
\bibitem{nlc1} Christensen N, Dupuis R J, Woan G and Meyer R 2004 {\it gr-qc/0402038}
\bibitem{TierneyMira99} Tierney L and Mira A 1999 {\it Statistics in Medicine} {\bf 18} 2507
\bibitem{Kirk83} Kirkpatrick S, Gelatt C D and Vecchi M P 1983 {\it Science}
{\bf 4598} 671
\bibitem{DupuisWoan03} Dupuis R J and Woan G 2003
preprint, "A Bayesian method to search for periodic gravitational waves"
\bibitem{Taylor} Taylor J H 2002 {\it Phys. Rev. D} {\bf 66} 084025
\bibitem{GreenMira01} Green P J and Mira A 2001 {\it Biometrika} {\bf 88} 1035
\bibitem{Mira98} Mira A 1998 {\it Ordering, Slicing and Splitting Monte Carlo Markov chain}
PhD thesis, Univertity of Minnesota

\endbib
\end{document}